\pdfoutput=1
\documentclass[doublecol]{epl2}
\usepackage{graphicx,color}
\usepackage{amsmath,amssymb}

\newcommand{\eref}[1]{eq.~(\ref{#1})}%
\newcommand{\Eref}[1]{Equation~(\ref{#1})}%
\newcommand{\fref}[1]{fig.~\ref{#1}} %
\newcommand{\Fref}[1]{Figure~\ref{#1}}%

\newcommand{\sgn}[1]{\mathrm{sgn}({#1})}%
\newcommand{\erfc}{\mathrm{erfc}}%

\begin{document}

\title{Work fluctuations for a harmonic oscillator  driven by an
external random force}

\author{Sanjib Sabhapandit}

\institute{Raman Research Institute, Bangalore 560080, India}

\date{\today}

{Fluctuation phenomena, random processes, noise, and Brownian motion} 
{Nonequilibrium and irreversible thermodynamics}

\abstract{
The fluctuations of the work done by an external Gaussian random force
on a harmonic oscillator that is also in contact with a thermal bath
is studied.  We have obtained the exact large deviation function as
well as the complete asymptotic forms of the probability density
function. The distribution of the work done are found to be
non-Gaussian. The steady state fluctuation theorem holds only if the
ratio of the variances, of the external random forcing and the thermal
noise respectively, is less than 1/3. On the other hand, the transient
fluctuation theorem holds (asymptotically) for all the values of that
ratio. The theoretical asymptotic forms of the probability density
function are in very good agreement with the numerics as well as with
an experiment.}

\maketitle

One of the most fundamental and important problems in the
nonequilibrium physics is to understand fluctuations. In this context,
the so-called \emph{fluctuation theorem} (FT) has generated a lot of
interest.  The FT was found first for the phase space contraction in
dynamical systems~\cite{Evans:93, Gallavotti:95} and later for a
certain ``action functional'' in stochastic systems~\cite{Kurchan:98,
Lebowitz:99} --- these quantities are generally referred to as the
``entropy production''.  Subsequently, there has been an increased
interest in the FTs for various physical quantities such as work,
power flux, heat flow, total entropy, etc.~\cite{Farago:02, vanZon:03,
Mazonka:99} --- because, in the absence of a general framework for
nonequilibrium phenomena, the FTs seem to be providing an unifying
picture for a variety of nonequilibrium systems. The so-called
Jarzynski equality~\cite{Jarzynski:97}, Crooks
relation~\cite{Crooks:99}, and Hatano-Sasa identity~\cite{Hatano:01}
are closely related to the FT.  In the linear response regime, the FT
leads to the Green-Kubo formula and the Onsager reciprocity
relations~\cite{Lebowitz:99, Gallavotti:96}.  However, the FT is more
general, as it also describes fluctuations in the nonlinear regime
arbitrarily far from the equilibrium.

The FT relates the positive and the negative fluctuations of a certain
time-integrated physical quantity $W_\tau=\int_0^\tau \dot{W}(t)\,
dt$, during a nonequilibrium process, according to:
\begin{equation}
\lim_{\tau\rightarrow\infty}\frac{1}{\tau}
\ln\biggl[ \frac{P(W_\tau=w\tau)}{P(W_\tau=-w\tau)}\biggr] = w,
\label{FT}
\end{equation}
where $P(W_\tau=\pm w\tau)$ is the probability density function (PDF)
of the physical quantity $W_\tau$ to have a value $\pm w\tau$.  In
fact, depending on the choice of the initial ensemble, there are two
kinds of FTs: the \emph{transient fluctuation theorem} (TFT) --- in
which the system at $\tau=0$ is in equilibrium, and the \emph{steady
state fluctuation theorem} (SSFT) --- in which the quantity $W_\tau$
is computed in a time interval $\tau$ in the nonequilibrium steady
state. Usually, the TFT is stated for a finite $\tau$, i.e., without
the limit $\tau\rightarrow\infty$ in \eref{FT}.  Naively, one would
expect the TFT and the SSFT to become equivalent in the
$\tau\rightarrow\infty$ limit. However, this is not always correct.

There have been several experimental tests of the FT and related
results, in diverse systems such as a colloidal particle in a changing
optical trap~\cite{Wang:02, Wang:05, Carberry:04}, liquid crystal
electroconvection~\cite{Goldburg:01}, fluidized granular
medium~\cite{Feitosa:04}, electrical circuits~\cite{Garnier:05}, RNA
stretching~\cite{Liphardt:02, Collin:05}, sheared micellar
gel~\cite{Majumdar:08}, harmonic oscillator~\cite{Douarche:06},
self-propelled polar particle~\cite{Kumar:10}, wave
Turbulence~\cite{Falcon:08}, and a gravitational wave
detector~\cite{Bonaldi:09}. A recent review of the experimental
applications of the FTs may be found in ref.~\cite{Ciliberto:10-2}.
Interpretation of experimental findings are not always easy as the FTs
are governed by the \textit{atypical} fluctuations that correspond to
the tails of the probability distributions --- and in an experiment in
a finite time, it is often hard to acquire enough of the rare events
to produce the tail of the distribution accurately.  Therefore, it is
very important to have exact theoretical predictions.

Theoretical investigations of the work FTs so far have been mostly
limited to the systems describe by linear Langevin equations with a
Gaussian white thermal noise and driven out of equilibrium by an
external \emph{deterministic} force.  In such cases~\cite{vanZon:03},
the distributions of the work done by the external force are Gaussian
and hence the work FTs hold somewhat trivially.  On the contrary, the
distributions of the work done by an external
Gaussian \emph{stochastic} force have been found to be non-Gaussian in
recent experiments on systems coupled to a thermal bath and driven out
of equilibrium by an external random force~\cite{Ciliberto:10}.
Motivated by these experiments, in this Letter, we address the
important question regarding the role of the external stochastic
forcing on the work fluctuations.

We consider one of the most basic physical systems, namely, the
harmonic oscillator. We investigate the fluctuations of the work done
by an externally applied Gaussian random force on a harmonic
oscillator that is also in contact with a thermal bath.  The
displacement $x(t)$ of the harmonic oscillator from its mean position
is described by the Langevin equation
\begin{equation}
m\frac{d^2x}{dt^2}+\gamma \frac{dx}{dt} +k x =\zeta_T (t) +f_0 (t),
\label{Langevin}
\end{equation}
where $m$ is the mass, $\gamma$ is the viscous drag coefficient and
$k$ is the spring constant.  The interaction with the thermal bath is
modeled by a Gaussian white noise $\zeta_T(t)$ with zero-mean
$\langle \zeta_T(t)\rangle=0$.  The externally applied force $f_0(t)$
is again a Gaussian random variable with $\langle f_0(t) \rangle =0$,
and $\zeta_T$ and $f_0$ are uncorrelated.  \Eref{Langevin} is
asymmetric in $\zeta_T$ and $f_0$ --- the fluctuation-dissipation
theorem relates the thermal fluctuation to the viscous drag as
$\langle \zeta_T(s) \zeta_T(t) \rangle= 2 D \delta(s-t)$ where
$D=\gamma k_B T$ with $T$ being the temperature of the bath and $k_B$
being the Boltzmann constant, whereas the fluctuation of the external
force $\langle f_0(s) f_0(t) \rangle = (\delta f_0)^2 \delta (s-t)$ is
independent of $\gamma$.  As it turns out, the only relevant parameter
is
\begin{equation}
\alpha=
\frac{(\delta f_0)^2}{2D}=
\frac{\langle x^2\rangle}{\langle x^2\rangle_\text{eq}}-1, \quad\text{and}~
\alpha \in (0,\infty),
\end{equation}
where $\langle x^2\rangle$ and $\langle x^2\rangle_\text{eq}$ are the
variance of $x$ in the steady-state (for $f_0 \not= 0$) and in
equilibrium (for $f_0=0$) respectively.

The quantity of interest is the work done by the external random force
$f_0(t)$ on the harmonic oscillator in a time interval $\tau$, in the
nonequilibrium steady state. This is given (in units of $k_B T$) by
\begin{equation}
W_\tau=\frac{1}{k_B T}\int_0^\tau  f_0(t) \frac{dx}{dt} \, dt,
\label{Work}
\end{equation}
with the initial condition (at $\tau = 0$) drawn from the steady state
distribution.  Evidently, $W_\tau$ is a fluctuating quantity whose
value depends on the initial condition, the trajectories of thermal
noise $\{\zeta_T(t): 0\le t \le \tau\}$ and the external random force
$\{f_0(t): 0\le t \le \tau\}$, during any particular realization.

It is clear from \eref{Langevin} that both the displacement $x$ and
the velocity $v=dx/dt$ depend linearly on the thermal noise and the
external random force. Therefore, the distribution of the phase space
variables $(x,v)$ is a Gaussian whose covariance matrix can be easily
evaluated from \eref{Langevin}. However, due to the nonlinear
dependence of the work given by \eref{Work}, on the thermal noise and
the external random forcing, the PDF $P(W_\tau)$ is not expected to be
Gaussian --- although for any fixed realizations of $\{ f_0(t)\}$ the
work fluctuation would be Gaussian.  Nonetheless, one expects the
large deviation form~\cite{Touchette:09}
\begin{equation}
P(W_\tau=w\tau/\tau_\gamma)\sim e^{(\tau/\tau_\gamma)\, h(w)}
\quad\text{for}~ \tau \gg \tau_\gamma,
\label{LDform}
\end{equation}
where $\tau_\gamma=m/\gamma$ is the viscous relaxation time and $h(w)$
is the large deviation function (LDF), which is defined by
\begin{equation}
h(w)=\lim_{(\tau/\tau_\gamma)\rightarrow\infty}\frac{1}{(\tau/\tau_\gamma)} \ln P(W_\tau=w\tau/\tau_\gamma).
\label{LDF}
\end{equation}
The FT
as given by ~\eref{FT} is equivalent to the symmetry relation
\begin{equation}
h(w)-h(-w)=w.
\label{symmetry}
\end{equation}
Our aim is to obtain the LDF $h(w)$ exactly, as well as the complete
asymptotic form of the PDF $P(W_\tau)$.

We begin by considering the characteristic function
\begin{equation}
\langle e^{-\lambda W_\tau} \rangle\equiv \int_{-\infty}^\infty dW_\tau\,
e^{-\lambda W_\tau} P(W_\tau)=Z(\lambda,\tau),
\label{Z}
\end{equation}
where $\langle \cdots\rangle$ denotes an average over the histories of
the thermal noise and the random forcing as well as the initial
condition.  The restricted characteristic function
$Z(\lambda,x,v,\tau|x_0,v_0)$ --- where the expectation is taken over
all trajectories of the system that evolve from a given initial
configuration $(x_0,v_0)$ to a given final configuration $(x,v)$ in
time $\tau$ --- satisfies the Fokker-Planck equation
$\bigl[\partial_\tau -\mathcal{L}_\lambda\bigr]\,
Z(\lambda,x,v,\tau|x_0,v_0)=0$ with the initial condition
$Z(\lambda,x,v,0|x_0,v_0)= \delta(x-x_0)\delta(v-v_0)$, where the
Fokker-Planck operator is given by
\begin{multline}
\mathcal{L}_\lambda=(1+\alpha)\frac{D}{m^2}\frac{\partial^2}{\partial^2v}
+
\biggl[\frac{k}{m} x + \frac{\gamma}{m}(1+2\alpha\lambda)v \biggr]
\frac{\partial}{\partial v} \\
-v\frac{\partial}{\partial x}
+\frac{\alpha\lambda^2\gamma^2}{D} v^2
+\frac{\gamma}{m}(1+\alpha\lambda).
\label{FP-op}
\end{multline}
The solution of the Fokker-Planck equation can be formally expressed
in the eigenbases of the operator $\mathcal{L}_\lambda$ and the large
$\tau$ behavior is dominated by the term having the largest
eigenvalue. Thus, for large $\tau$,
\begin{equation}
  Z(\lambda,x,v,\tau |x_0,v_0) 
\sim
  \chi(x_0,v_0,\lambda)\Psi(x,v,\lambda)\,
  e^{\tau\mu(\lambda)},
  \label{characteristic.1}
\end{equation}
where $\Psi(x,v,\lambda)$ is the eigenfunction corresponding to the
largest eigenvalue $\mu(\lambda)$ and $\chi(x_0,v_0,\lambda)$ is the
projection of the initial state onto the eigenstate corresponding to
the eigenvalue $\mu(\lambda)$.  To calculate these functions, we
follow an approach that was used recently to compute the fluctuations
of the heat transport across a harmonic chain~\cite{Kundu:11}.
Skipping details~\cite{Details}, we find that
\begin{align}
\label{mu}
&\mu(\lambda)=\frac{1}{2\tau_\gamma}\bigl[1-\eta(\lambda) \bigr], 
~~ \eta(\lambda)=\sqrt{1+4\alpha\lambda(1-\lambda)},\\
\label{psi}
&\Psi(x,v,\lambda)=\left[\frac{\gamma\eta(\lambda)\sqrt{km}}
{2\pi(1+\alpha)D}\right]
\exp\bigl[-B_+(\lambda) E(x,v)\bigr],
\\
\label{chi}
&\text{and}\quad\chi(x_0,v_0,\lambda)= 
\exp\bigl[-B_-(\lambda) E(x_0,v_0)\bigr],
\end{align}
where
\begin{equation}
B_\pm(\lambda)=
\frac{\gamma \bigl[\eta(\lambda)\pm (1 + 2 \alpha \lambda)\bigr]}{
2 (1+\alpha)D},
\end{equation}
and 
\begin{equation}
E(x,v)=\frac{1}{2} k x^2 + \frac{1}{2} m v^2,
\end{equation}
is the total energy of the harmonic oscillator.  Note from \eref{mu}
that the largest eigenvalue satisfies the symmetry relation
$\mu(\lambda)=\mu(1-\lambda)$, even though $\mathcal{L}_\lambda$ and
its adjoint $\mathcal{L}_\lambda^\dagger$ do not possess the symmetry
$\mathcal{L}_\lambda^\dagger = \mathcal{L}_{1-\lambda}$.

Using the explicit forms of eqs.~\eqref{FP-op}
and \eqref{mu}--\eqref{chi}, the eigenvalue equation
$\mathcal{L}_\lambda \Psi (x,v,\lambda)=\mu(\lambda)\Psi(x,v,\lambda)$
and the normalization $\int_{-\infty}^\infty\int_{-\infty}^\infty
\chi(x,v,\lambda) \Psi(x,v,\lambda)\, dx\, dv =1$
can be indeed verified.  Moreover, $\mu(0)=0$ and $\chi(x_0,v_0,0)=1$,
which is expected --- since \eref{FP-op} for $\lambda=0$, corresponds
to the Fokker-Planck operator of the phase space variables, and hence
the steady state distribution $Z(\lambda=0,x,v,\tau\rightarrow\infty
|x_0,v_0)$ must be independent of the initial condition and $\tau$.
The steady state distribution of $(x,v)$ is given by
$Z(\lambda=0,x,v,\tau\rightarrow\infty |x_0,v_0)=\Psi(x,v,0)$.

Now, substituting Eqs.~\eqref{psi} and \eqref{chi}
in \eref{characteristic.1}, then averaging over the initial variables
$(x_0,v_0)$ with respect to $\Psi(x_0,v_0,0)$ and integrating over the
final variables $(x,v)$, we find the characteristic function that is
defined by \eref{Z}, as
\begin{equation}
Z(\lambda,\tau) \sim g(\lambda)\,   e^{\tau\mu(\lambda)},
\label{Z-asymptotic}
\end{equation}
where $\mu(\lambda)$ is given by \eref{mu} and 
\begin{equation}
g(\lambda)=
\frac{2}
{1+\eta(\lambda) -2\alpha\lambda}
\times
\frac{2\eta(\lambda)}
{1+\eta(\lambda) +2\alpha\lambda}~.
\label{g(lambda)}
\end{equation}
The first factor in the above equation is due to the averaging over
the initial conditions with respect to the the steady state
distribution and the second factor is due to the integrating out of
the final degrees of freedom.

The PDF of the work done is related to its characteristic function by
the inverse Fourier transform
\begin{equation}
P(W_\tau) = \frac{1}{2\pi i} \int_{-i\infty}^{+i\infty} Z(\lambda,\tau)\,
e^{\lambda W_\tau}\, d\lambda,
\end{equation}
where the integration is done along the imaginary axis (vertical
contour through the origin) in the complex $\lambda$ plane.  The large
$\tau$ ($\gg \tau_\gamma$) behavior of $P(W_\tau)$ can be obtained
from the saddle point approximation of the above integral while using
the asymptotic form of $Z(\lambda,\tau)$ given by \eref{Z-asymptotic}.
We note that $\eta(\lambda)$, given in \eref{mu}, has two branch
points on the real $\lambda$ line at
\begin{equation}
\lambda_\pm=\frac{1}{2}\left[1\pm \sqrt{1+\frac{1}{\alpha}} \right], 
\label{lambda_pm}
\end{equation}
as
$\eta(\lambda)=\sqrt{4\alpha(\lambda_+-\lambda)(\lambda-\lambda_-)}$~. Outside
the interval $[\lambda_-,\lambda_+]$ on the real $\lambda$ line,
$\eta(\lambda)$ is imaginary. However, $Z(\lambda,\tau)$ must be real
for real values of $\lambda$, if the integral in \eref{Z} converges.
Therefore, analytical continuation of $Z(\lambda)$ to the real
$\lambda$ is allowed only within the range $\lambda_- < \lambda
< \lambda_+$ --- for which $[\eta(\lambda)]^2 > 0$, and hence,
$\mu(\lambda)$ is real and analytic. In fact, in the whole complex
$\lambda$ plane, $\eta(\lambda)$ is real only for $\lambda$ in the
real interval $[\lambda_-,\lambda_+]$. Therefore, we expect the saddle
to be also in that interval.

Now, in the expression of $g(\lambda)$ given by \eref{g(lambda)}, the
denominator of the second factor is positive for
$\lambda\in(\lambda_-,\lambda_+)$ for all $\alpha \in
(0,\infty)$. Hence, the second factor of $g(\lambda)$ is analytic in
the interval $(\lambda_-,\lambda_+)$. On the other hand, the analytic
properties of the first factor in \eref{g(lambda)}, depends on the
value of the parameter $\alpha$.

As long as $\alpha < 1/3$, the denominator of the first factor is
positive for $\lambda\in(\lambda_-,\lambda_+)$. Therefore, in this
case $g(\lambda)$ is analytic in $(\lambda_-,\lambda_+)$ and hence can
be neglected in the saddle-point calculation as a subleading
contribution. The saddle-point calculation with $Z(\lambda,\tau)\sim
e^{\tau\mu(\lambda)}$ relates $\mu(\lambda)$ to the LDF $h(w)$
of \eref{LDform}, by the Legendre transform
\begin{equation}
h(w)=\tau_\gamma \mu(\lambda^*) +\lambda^* w, 
\qquad
-\tau_\gamma\mu'(\lambda^*)=w.
\end{equation}
In this case, the symmetry relation of the LDF as given
by \eref{symmetry}, follows directly from the symmetry
$\mu(\lambda)=\mu(1-\lambda)$. The solution of the condition
$-\tau_\gamma \mu'(\lambda^*)=w$ gives the saddle point $\lambda^*$ in terms of
$w$ as
\begin{equation}
\lambda^*(w)=\frac{1}{2}
\left[1-\frac{w}{\sqrt{w^2+\alpha}}\sqrt{1+\frac{1}{\alpha}} \right].
\label{lambda*}
\end{equation}

We now consider the case $\alpha > 1/3$.  In this case, due to the
first factor in \eref{g(lambda)}, $g(\lambda)$ possesses a pole at
\begin{equation}
\lambda_0=\frac{2}{1+\alpha},
\end{equation}
and $\lambda_- < 0 <\lambda_0 < \lambda_+ $.  Now, $g(\lambda)$ is
negative for $\lambda>\lambda_0$. However, $g(\lambda)$ must be
non-negative for any real $\lambda$, if the integral in \eref{Z}
exists. Therefore, now the allowed range of real $\lambda$ shrinks to
$(\lambda_-,\lambda_0)$.  It follows from \eref{lambda*} that
$\lambda^*(w)$ is a monotonically decreasing function of $w$, and
$\lambda^*(w\rightarrow\mp \infty)\rightarrow\lambda_\pm$.  Note that
$\lambda^*\in (\lambda_-,\lambda_+)$ as expected.  For any given
$\alpha$, as $w$ decreases from $+\infty$ to $-\infty$, the saddle
point $\lambda^*(w)$ moves unidirectionally from $\lambda_-$ to
$\lambda_+$.  Thus, for sufficiently large $w$, we have $\lambda_-
< \lambda^* <\lambda_0$. In such situation, the contour of integration
can be deformed smoothly through the saddle point $\lambda^*$, and
therefore, the LDF is still given by $h(w)=\tau_\gamma \mu(\lambda^*)
+\lambda^* w$.  However, as one decreases $w$, at some particular
value $w=w^*$, the saddle-point hits the singularity.  For $w<w^*$, we
then have $0 < \lambda_0 <\lambda^*$.  In this case, the leading
contribution comes essentially from the pole~\cite{Details}, which
yields $h(w)= \tau_\gamma \mu(\lambda_0) +\lambda_0 w$.  Using
$\lambda^*(w^*)=\lambda_0$ and $\mu'(\lambda^*) + w =0$, it is easy to
check that $h(w)$ and its derivative $h'(w)$ are continuous at
$w=w^*$.  For $\alpha = 1/3$, we have
$\lambda_0=\lambda_+=3/2$. Since, $\lambda^*\rightarrow \lambda_+$,
only when $w\rightarrow-\infty$, for any finite $w$ we again have
$h(w)=\tau_\gamma \mu(\lambda^*) +\lambda^* w$, i.e., $w^*=-\infty$.

\begin{figure}
\includegraphics[width=.9\hsize]{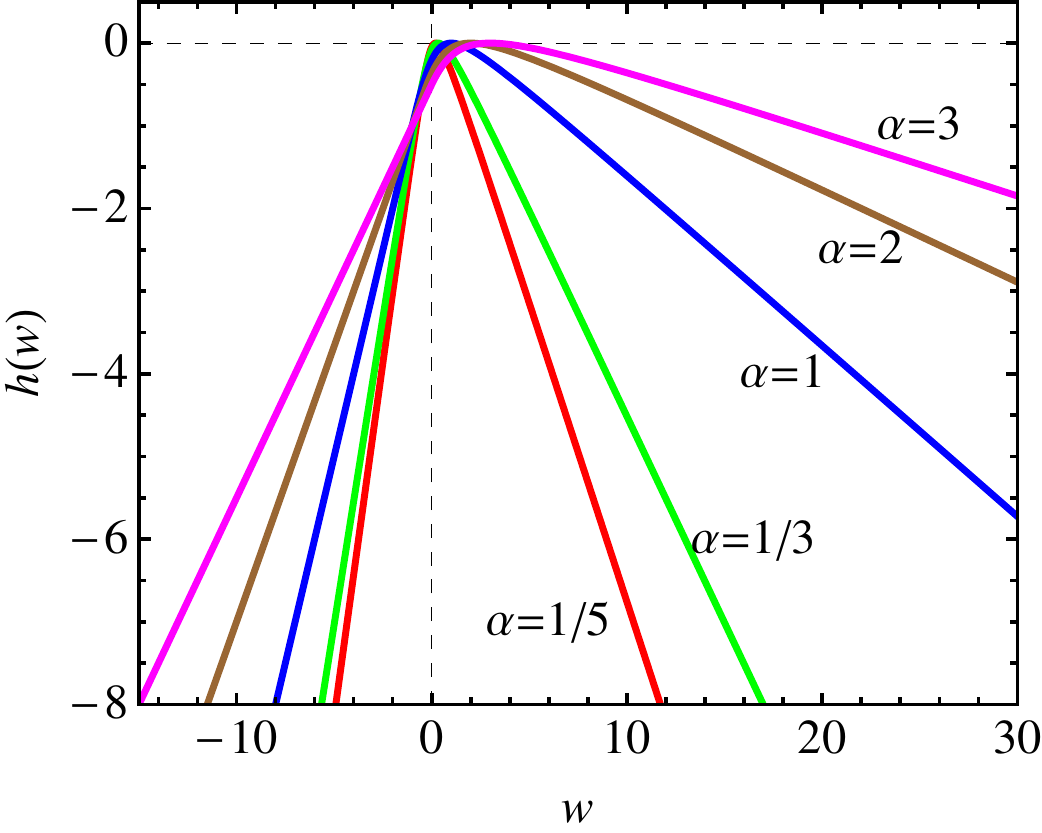} 
\caption{\label{h-figure}(Color
 online). Plot of the large deviation function $h(w)$ for $\alpha=1/5$
 (red), $1/3$ (green), $1$ (blue), $2$ (brown), and $3$ (magenta).}
\end{figure}

Let us express the LDF $h(w)$, defined by \eref{LDF}, explicitly in
terms of $w$ and $\alpha$.
We find that, for $\alpha \le 1/3$:
\begin{equation}
h(w)=h_1(w),
\label{h(w)-1}
\end{equation}
and 
for $\alpha \ge 1/3$:
\begin{equation}
h(w)=
\begin{cases}
h_1(w)
 & \text{for}~ w \ge w^*\\
h_2(w) & \text{for}~ w \le w^*
\end{cases},
\label{h(w)-2}
\end{equation}
where $h_1(w)$ and $h_2(w)$ are given by
\begin{align}
h_1(w)&=\frac{1}{2}\left[1+w
- \sqrt{w^2+\alpha}\sqrt{1+\frac{1}{\alpha} }\right],\\
h_2(w) &=\frac{1-\alpha}{1+\alpha} + \frac{2w}{1+\alpha},
\end{align}
and $w^*$ is found by solving $\lambda^*(w^*)=\lambda_0$, as
\begin{equation}
w^*=\frac{\alpha(\alpha-3)}{3\alpha-1}.
\label{w*}
\end{equation}
\Fref{h-figure} displays the LDF for various $\alpha$.

\begin{figure}
\includegraphics[width=.9\hsize]{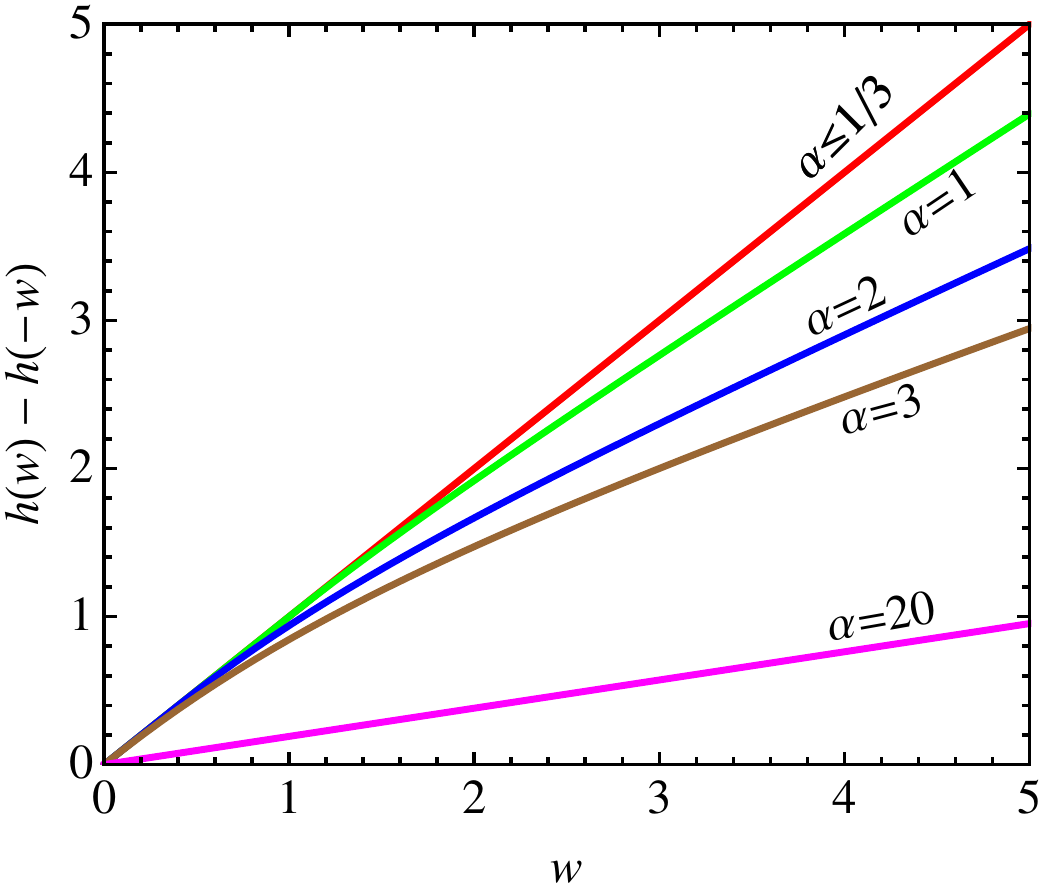} \caption{\label{FR-figure}(Color
 online).  Asymmetry function for various $\alpha$.}
\end{figure}

From the above expressions, it is now straightforward to check the
validity of the work SSFT.  For $\alpha \le 1/3$, we get
$h(w)-h(-w)=w$, which implies that the SSFT is satisfied.  On the
other hand, for $1/3 <\alpha <3$, we get $h(w)-h(-w)=w$ only for
$w^*<w<-w^*$. For $\alpha \ge 3$, the symmetry
relation \eqref{symmetry} is not satisfied for any $w$.  For example,
for $\alpha=3$, we get $w^*=0$ and $h(w)-h(-w)=1+w-\sqrt{1+(w^2/3)}$~.
\Fref{FR-figure} displays the asymmetry function 
$\rho(w)\equiv h(w)-h(-w)$ for several values of $\alpha$.

We can also obtain the the complete asymptotic form of the PDF of the
work fluctuations. Skipping details~\cite{Details}, we find that for
$\alpha \le 1/3$:
\begin{equation}
P(W_\tau=w\tau/\tau_\gamma) \approx \frac{K(w)}{2\sqrt{\pi (\tau/\tau_\gamma)}}\, 
e^{(\tau/\tau_\gamma) h_1(w)},
\label{PDF-1}
\end{equation}
and for $\alpha \ge 1/3$:
\begin{widetext}
\begin{align}
P(W_\tau=w\tau/\tau_\gamma) \approx\, &  \frac{ e^{(\tau/\tau_\gamma)
h_1(w)}}{2\sqrt{\pi (\tau/\tau_\gamma)}}\, \left[K(w)
-\frac{\sgn{w^*-w}\,g_{-1}}{\sqrt{h_2(w)-h_1(w)}}\right]
\notag\\[2mm]
&+  e^{(\tau/\tau_\gamma) h_2(w)}\, g_{-1} \left[
 \frac{\sgn{w^*-w}}{2}\, \erfc \Bigl(\sqrt{\tau[h_2(w)-h_1(w)]}\Bigr)
 -\theta(w^*-w)
\right].
\label{PDF-general}
\end{align}
\end{widetext}
\begin{floatequation} 
\mbox{\textit{see eq.~\eqref{PDF-general}}},
\end{floatequation}\addtocounter{equation}{-1}
where
\begin{align}
K(w)
&= \alpha^{3/2} (1+1/\alpha)^{3/4} (w^2+\alpha)^{-5/4} \notag\\
&\times \Bigl[1+(w+\alpha) \lambda^*(w) - h_1(w)\Bigr]^{-1}\notag\\
&\times \Bigl[\bigl\{h_2(w)-h_1(w)\bigr\} +
(w-\alpha) \bigl\{\lambda^*(w) -\lambda_0\bigr\}\Bigr]^{-1},
\label{K(w)}
\end{align}
and
\begin{equation}
g_{-1}
=\lim_{\lambda\rightarrow\lambda_0} \bigl[(\lambda-\lambda_0)\,
g(\lambda) \bigr]
=-\frac{(3\alpha-1)^2}{8\alpha^2(1+\alpha)}~.
\label{g-1}
\end{equation}

\begin{figure*}
\begin{center}
\includegraphics[width=.32\hsize]{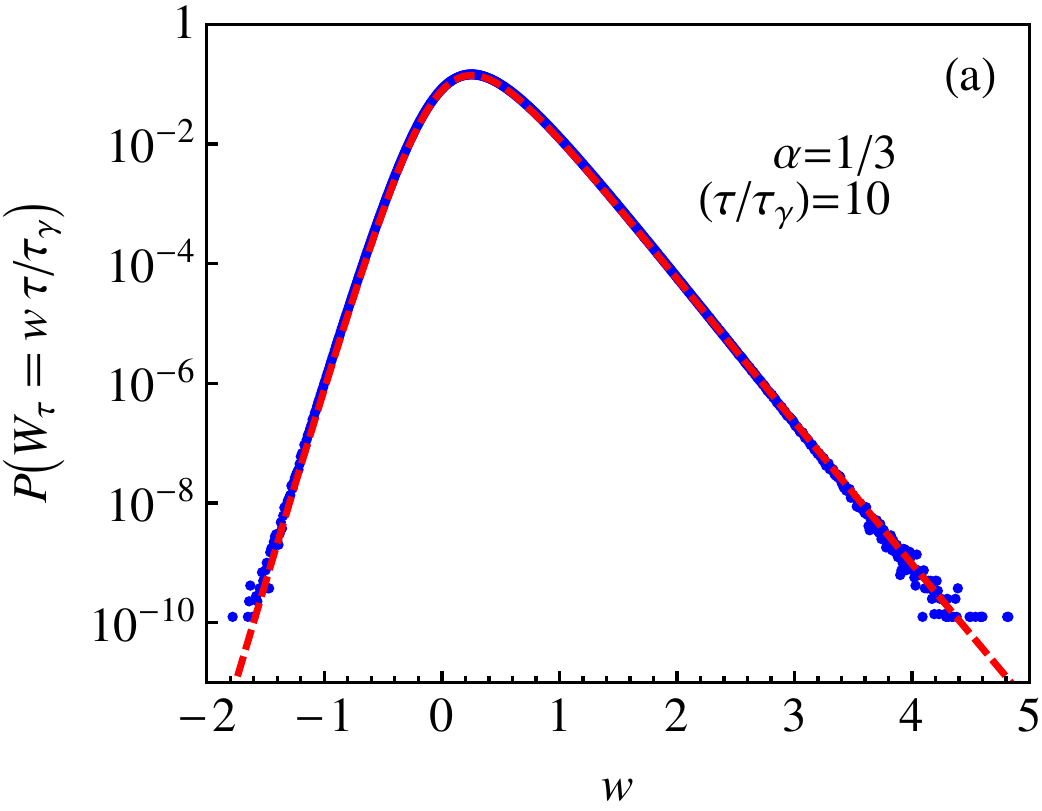} 
~~\includegraphics[width=.32\hsize]{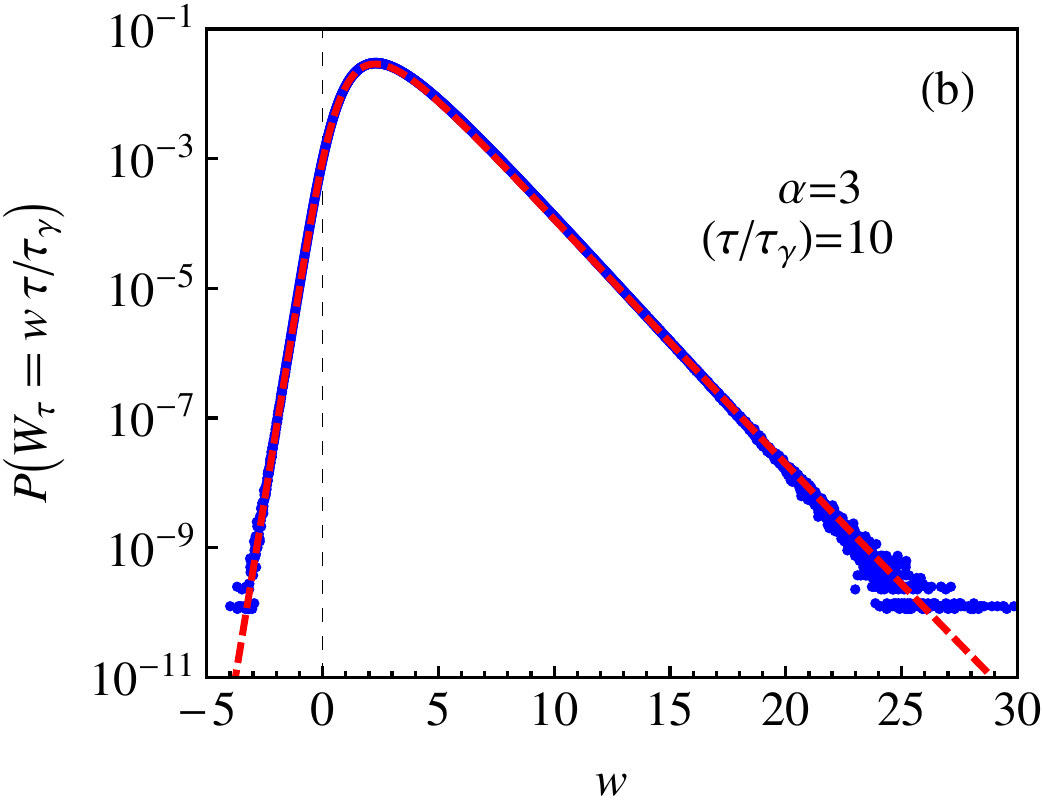}
~~\includegraphics[width=.32\hsize]{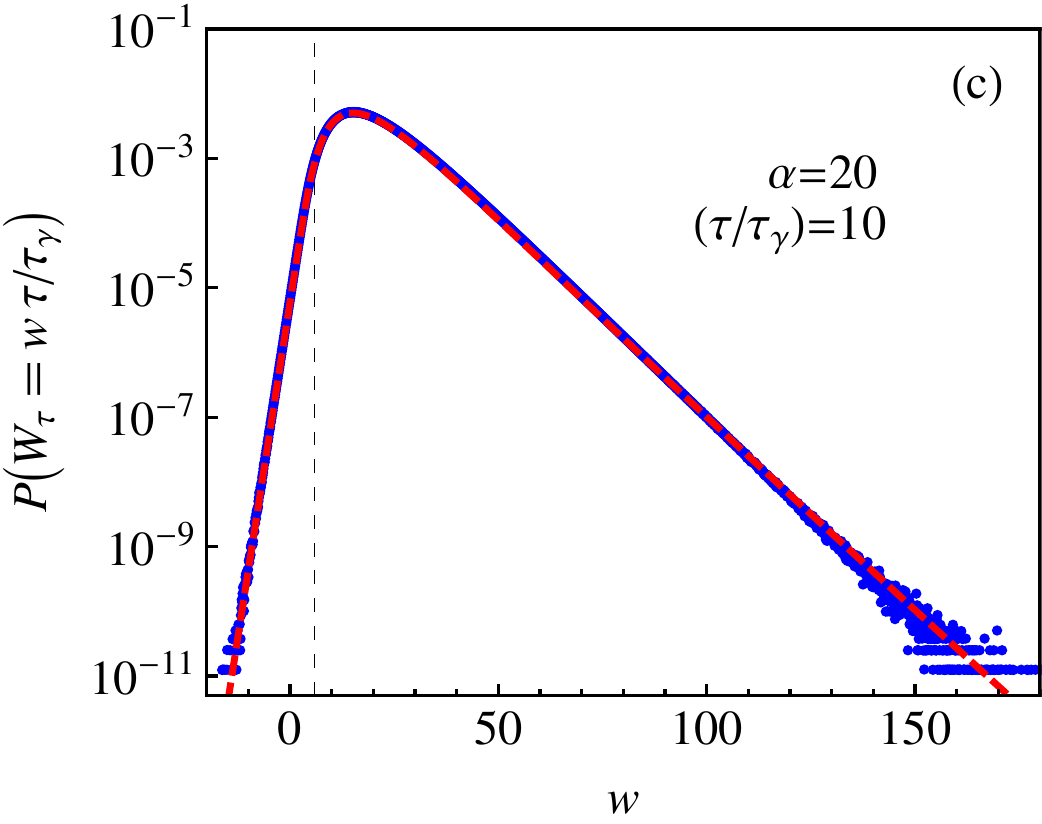}
\end{center}
\caption{\label{PDF-figure}(Color online). $P(W_\tau)$ against the scaled variable $w=(\tau/\tau_\gamma)^{-1}W_\tau$ for $(\tau/\tau_\gamma)=10$ and (a) $\alpha=1/3$, (b) $\alpha=3$ and (c)
 $\alpha=20$ respectively. The points (blue) are obtained from
 numerical simulation, and the thick dashed lines (red) plot the
 analytical asymptotic forms given by \eref{PDF-1} for (a)
 and \eref{PDF-general} for (b) and (c). The vertical thin dashed
 lines in (b) and (c) mark the position of $w=w^*$.}
\end{figure*}

We compare the above asymptotic forms of the PDF with the results
obtained from the numerical simulations of the Langevin
equation \eqref{Langevin}. As seen from \fref{PDF-figure}, the
agreements are extremely good, even for $\tau/\tau_\gamma=10$.

\begin{figure*}
\begin{center}
\includegraphics[width=.45\hsize]{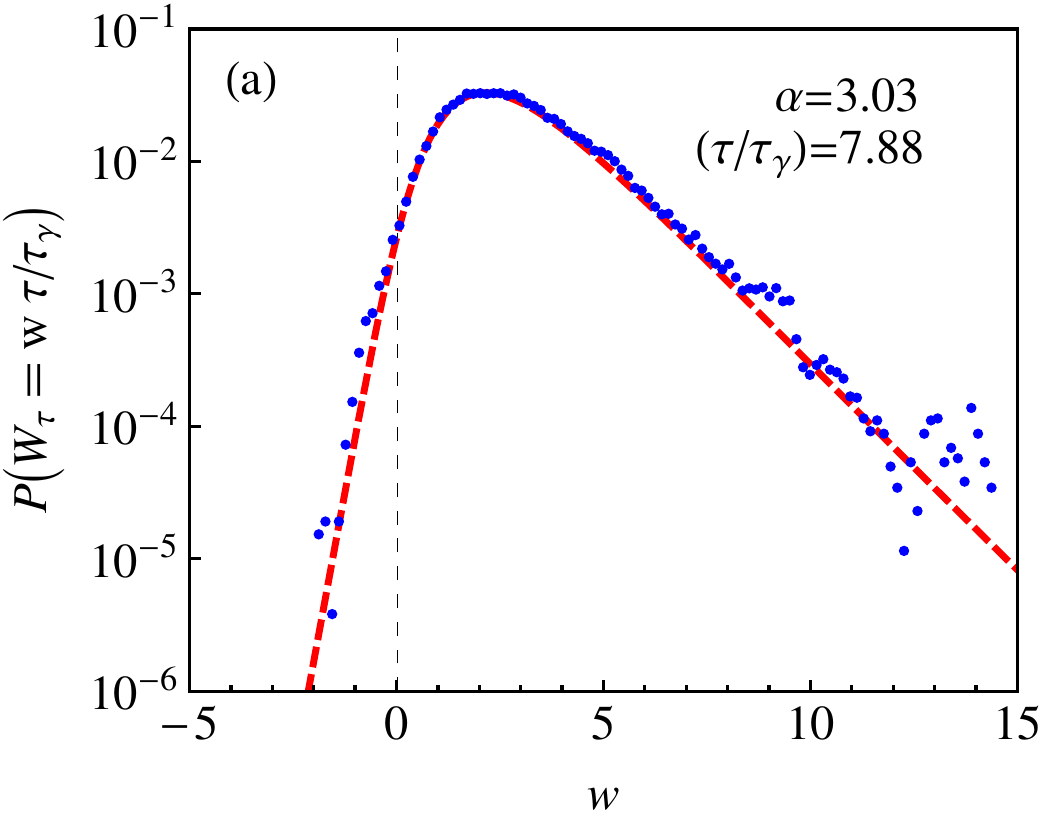}~~~~
\includegraphics[width=.45\hsize]{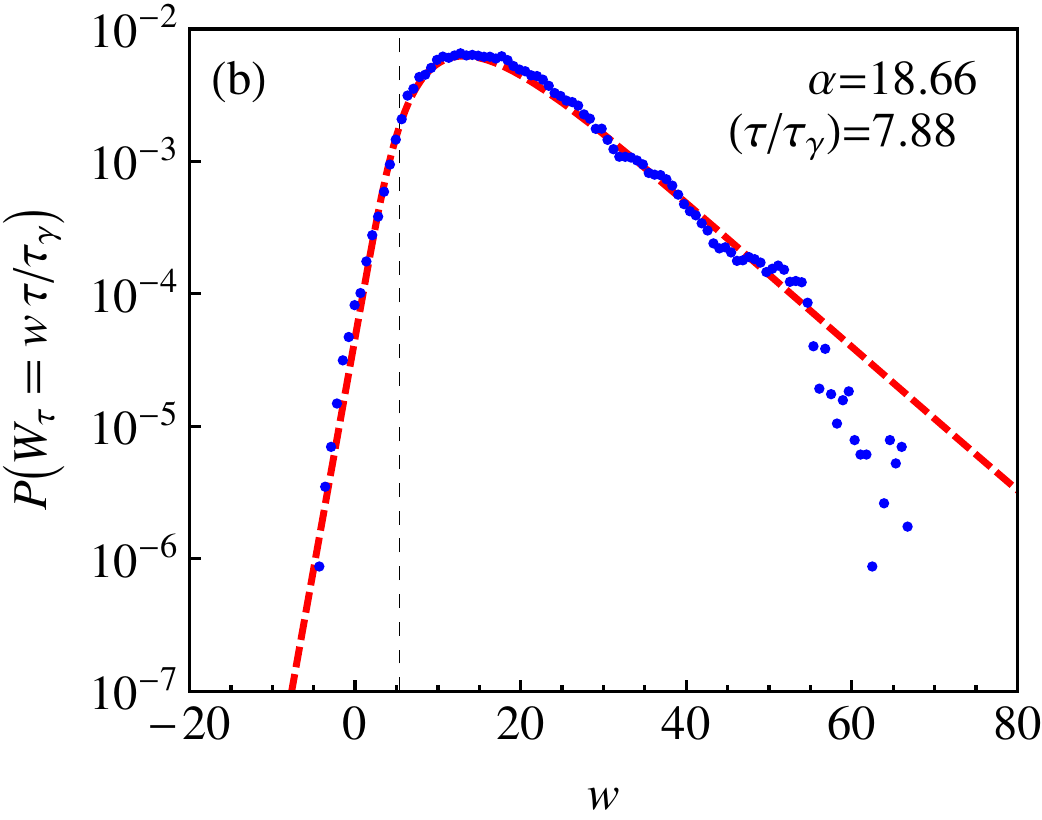}
\end{center}
\caption{\label{PDF-figure-expr}(Color online). $P(W_\tau)$ against
the scaled variable $w=(\tau/\tau_\gamma)^{-1} W_\tau $ for
$(\tau/\tau_\gamma)=7.88\dots$, and (a) $\alpha=3.03$ and (b)
$\alpha=18.66$ respectively. The points (blue) are from the
micro-cantilever experiment reported in ref.~\cite{Ciliberto:10}, and
the thick dashed lines (red) plot the analytical asymptotic form
given by \eref{PDF-general}. The vertical thin dashed lines mark the
position of $w=w^*$.}
\end{figure*}

We also compare our analytical results against an experiment that was
carried out on an atomic-force microscopy cantilever and reported in
ref.~\cite{Ciliberto:10}. The dynamics of the micro-cantilever tip is
described by \eref{Langevin} with the viscous relaxation time
$\tau_\gamma=632\,\mu s$.  We find very good agreement between the
theory and the experiment, which is shown in \fref{PDF-figure-expr}

So far, we have considered the fluctuations of $W_\tau$ in the
nonequilibrium steady state, as we have averaged over the initial
conditions in \eref{characteristic.1} with respect to the steady state
distribution $\Psi(x_0 , v_0 , 0)$ to arrive at
\eref{Z-asymptotic}. Let us now examine how the
nature of the initial state affects the results.  We recall that the
singular part of $g(\lambda)$, i.e., the first factor
in \eref{g(lambda)} comes from the averaging
of \eref{characteristic.1} with respect to the steady state
distribution of the initial state.  Without the averaging, for any
given initial configuration $(x_0,v_0)$, the resulting prefactor of
$e^{\tau\mu(\lambda)}$ remains analytic throughout the interval
$(\lambda_-,\lambda_+)$, and hence can be neglected from the saddle
point calculation as the subleading contribution.  Therefore, the FT
for a fixed initial condition is always satisfied, as the LDF is given
by \eref{h(w)-1} for all $\alpha \in (0,\infty)$.  If the initial
state at $\tau=0$ is chosen from equilibrium ---i.e., the average
in \eref{characteristic.1} is taken with respect to the Boltzmann
weight $\propto \exp[-E(x_0,v_0)/(k_B T)]$ --- then the first factor
in \eref{g(lambda)} is replaced by $2(1+\alpha)/[1+\eta(\lambda) +
2\alpha(1-\lambda)]$. In that case even $g(\lambda)$ satisfies the
symmetry relation $g(\lambda)=g(1-\lambda)$.  It is easy to see that,
now $g(\lambda)$ remains analytic in $(\lambda_-, \lambda_+)$ for any
$\alpha$. Therefore, the LDF in this case is again given
by \eref{h(w)-1} for all $\alpha \in (0,\infty)$. Consequently, the
TFT is satisfied (as $\tau\rightarrow \infty$) for all $\alpha \in
(0,\infty)$.

In conclusion, we have studied the work fluctuations of a harmonic
oscillator coupled to a thermal bath and driven out of equilibrium by
an external Gaussian random force.  We have found that the SSFT holds
only for weak forcing, whereas the TFT (with $\tau\rightarrow\infty$)
holds for all forcing.  More importantly, we have analytically
obtained the exact LDF as well as the complete asymptotic forms of the
PDF of the work fluctuations, and quite interestingly, they are
independent of the spring constant of the harmonic oscillator.
However, while the LDFs are same for both $k\not=0$ and $k=0$ cases,
the complete asymptotic form of the PDFs are different in the two
cases. Therefore, $k\rightarrow 0$ limit of PDF (which is anyway
independent of $k$) is not same as the PDF in the $k=0$ case.  The
nature of the work fluctuation is found to be non-Gaussian.  These
exact results should have broad and important applications, as the
harmonic oscillator is ubiquitous in nature.  For example, many
nanomechanical and biological systems are essentially described by a
harmonic oscillator and the results of this Letter are expected to be
useful there.

\acknowledgments
The author thanks Abhishek Dhar for useful discussions, and
J.~R.~Gomez-Solano and S.~Ciliberto for sending their data of the
experiment on cantilever reported in ref.~\cite{Ciliberto:10}.

\end{document}